\documentstyle[12pt]{article}
\newcommand{\be}{\begin{equation}}
\newcommand{\ee}{\end{equation}}
\def\la{\mathrel{\mathpalette\fun <}}
\def\ga{\mathrel{\mathpalette\fun >}}
\def\fun#1#2{\lower3.6pt\vbox{\baselineskip0pt\lineskip.9pt
\ialign{$\mathsurround=0pt#1\hfil##\hfil$\crcr#2\crcr\sim\crcr}}}

\title{\begin{flushright}
{\normalsize FAU-TP3-98/14\\
NUC-MINN-98/5-T\\
July 1998 \\
}
\end{flushright}
\vspace*{0.3in}
{\bf Mass Shift and Width Broadening of $\rho$-mesons
Produced in Heavy Ion Collisions}}
\author{{\bf V. L. Eletsky}$^{1,2}$, {\bf B. L. Ioffe}$^1$
 \vspace*{0.2in} and {\bf J. I. Kapusta}$^3$\\
   {$^1$\it Institute of Theoretical and Experimental Physics}\\
   {\it B. Cheremushkinskaya 25}\\  \vspace*{0.2in}
   {\it Moscow 117218, Russia}\\
   {$^2$\it Institut f\"ur Theoretische Physik III}\\
   {\it Universit\"at Erlangen-N\"urnberg}\\  \vspace*{0.2in}
   {\it D-91058 Erlangen, Deutschland}\\
   {$^3$\it School of Physics and Astronomy}\\
   {\it University of Minnesota}\\
   {\it Minneapolis, MN 55455, USA}}

\date{}

\begin{document}

\maketitle
\begin{abstract}

The mass shift $\Delta m_{\rho}$ and width broadening
$\Delta\Gamma_{\rho}$ of $\rho$-mesons produced in heavy ion collisions is
estimated using general formule which relate the in-medium mass shift of
a particle to the real part of the forward scattering amplitude ${\rm Re}
f(E)$ of this particle on constituents of the medium and $\Delta\Gamma$ to
the corresponding cross section.
It is found that the mass increases by some tens of MeV but, more
importantly, the width becomes large, increasing by several
hundred MeV at beam energies of a few GeV$\cdot$A and by twice that
amount at beam energies of about a hundred GeV$\cdot$A.

\end{abstract}

\newpage

\section{Introduction.}

The problem of how the properties of hadrons change in hadronic or
nuclear matter in comparison to their free values has attracted a lot
of attention recently. Among these properties of immediate interest
are the in-medium particles mass shifts and width broadenings.
Different models, as well as model independent
approaches, were used to calculate these effects both at finite
temperature and finite density.  For a review see \cite{ab}.  It is
clear on physical grounds that the in-medium mass shift and width
broadening of a particle is only due to its interaction with the
constituents of the medium.  Thus one can use phenomenological information
on this interaction to calculate the mass shifts. In a recent
paper\cite{ei}, two of us have argued that the mass
shift of a particle in medium can be related to the forward scattering
amplitude $f(E)$ of this particle on the constituents of the medium.
Written in the medium rest frame it is:

\be
\Delta m (E) = -2\pi\frac{\rho}{m}{\rm Re} f(E) \, .
\label{dm}
\ee
Here $m$ is the vacuum mass of the particle, $E$ is its energy in the rest
frame of the constituent particle, and $\rho$ is the density of consituents.
The normalization of the amplitude corresponds to the standard form of the
optical theorem,

\be
k\sigma = 4\pi {\rm Im} f(E) \, ,
\label{opt}
\ee
where $k$ is the particle momentum. The width broadening is given by
\be
\Delta \Gamma(E) = \frac{\rho}{m}~k\sigma(E) \, .
\ee
The domain of applicability of Eqs. (1) and (3) was discussed in
\cite{ei}.  Briefly:\\
\noindent
$\bullet$ The particle's wavelength $\lambda$ must be much less than the mean
distances between medium constituents $d$: $\lambda=k^{-1}\ll d$. This means
that the particle's momentum $k$ must be larger than a few hundred MeV.\\
\noindent
$\bullet$ The particle's formation length $l_f\sim (E/m)/m_{\rm char}$,
with $m_{\rm char}\approx m_{\rho}$, must be less than the nucleus radius $R$.\\
\noindent
$\bullet$ ${\rm Re}f(E)$, which enters Eq. (1), must satisfy the inequality
$\mid {\rm Re}f\mid < d$.\\
\noindent
$\bullet$ The main part of the scattering proceeds through small angles,
$\theta \ll 1$. Only in this case is the optical analogy
on which Eqs. (1) and (3) are based correct.

Equations (1) and (3) are correct also when the medium constituents
have some momentum distributions, such as Fermi-Dirac or Bose-Einstein
distributions for finite temperatures and chemical potentials.
In these cases averaging over the constituents' momentum distributions
must be performed on the right sides of Eqs. (1) and (3).
These equations were derived in [2] based on simple quantum
mechanical arguments and the optical analogy. This approach allows one to
formulate in an explicit way the applicability conditions presented above.
When the medium is a gas in thermal equilibrium the equivalent to
Eqs. (1) and (3) can be derived on the basis of thermal field theory.
References \cite{3,4} give a few examples and reference \cite{5} gives
a relativistic field-theoretic derivation.

In most of the papers on the in-medium hadron mass shifts the hadrons
were considered at rest.  As seen from Eq. (1) this restriction is not
necessary theoretically.  It is  desirable to have theoretical predictions
in a broad  energy range since it extends the possibilities of
experimental investigation.  As discussed in \cite{6} for the cases of
$\rho$ or $\pi$-mesons embedded in nuclear matter the energy dependence of
the mass shifts is rather significant at low energies where resonances
dominate.

We estimate the mass shift and width broadening in the
case of $\rho$-mesons produced in heavy ion collisions.
The most interesting case is that of the $\rho^0$ which can be observed
through the decay $\rho^0 \to e^+e^-$ or $\mu^+\mu^-$.  We will assume that
$\rho$-mesons are formed in the last stage of the evolution of hadronic
matter created in course of a heavy ion collision when the matter can be
considered as an almost noninteracting gas of pions and nucleons.
(We will neglect the admixture of kaons and hyperons, which is known
to be small \cite{7}, as well as heavy resonances.)
This stage occurs late in the collision when
the total density of nucleons and pions is of the
order of the normal nucleon density in a nucleus. The description of nuclear
matter as a noninteracting gas of nucleons and pions cannot be
considered as a very good one so it is clear from the beginning that our
results may be only semiquantitave. The main ingredients of our calculation
are $\rho\pi$ and $\rho N$ forward scattering amplitudes and total cross
sections as well as the values of nucleon and pion densities.

In this paper we consider central heavy ion collisions and assume that
nucleon and pion momentum distributions in the gas are just the momentum
distributions measured experimentally in such collisions. The case where
nucleons and pions are assumed to exist in a state of equilibrium at
fixed temperature and chemical potential will be considered in a subsequent
publication [8].

\section{Calculation of $\rho N$ and $\rho \pi$ cross sections
and forward scattering amplitudes.}

Let us first focus on the amplitudes and cross sections. To determine
these quantities we use the following procedure.  At low energies we
saturate the cross sections and forward scattering amplitudes by resonance
contributions. At high energies we determine $\sigma_{\rho N}$ and
$\sigma_{\rho \pi}$ from $\sigma_{\gamma N}$  and $\sigma_{\gamma\pi}$
using the vector dominance model (VDM).  The cross section
$\sigma_{\gamma N}$  is well known experimentally \cite{9},
${\rm Re}f_{\gamma N}$ is determined from the dispersion relation, and
$\sigma_{\gamma\pi}$ and ${\rm Re}f_{\gamma \pi}$ can be found by the Regge
approach.  Since VDM allows one to find only the cross sections
of transversally polarized $\rho$-mesons we restrict ourselves to this
case.  As was shown in [2], when $E_{\rho}\ga 2$ GeV, $\Delta m$ and $\Delta
\Gamma$ for longitudinal $\rho$-mesons are much smaller than for
transversal ones in nuclear matter.  At zero $\rho$-meson energy,
$\Delta m$ and $\Delta \Gamma$  for transverse and longitudinal
$\rho$-mesons are evidently equal.  In the case of scattering on
a low temperature pion gas they are comparable [10].
Therefore our results should be multiplied by a factor ranging
between 1 and 1.5 for unpolarized $\rho$-mesons.

To estimate ${\rm Re} f_{\rho\pi}(s)$ at low energy we write, in the center
of mass (c.m.) frame:
\be
{\rm Re} f_{\rho\pi}(s)=-\sum_{R} F_s F_i \frac{1}{2q_{cm}}
\frac{B_R\Gamma_R (\sqrt{s}-m_R)}{(\sqrt{s}-m_R)^2 +\Gamma_R^2/4} \, ,
\label{res}
\ee
where $\sqrt{s}$ is the total c.m. energy, $m_R$ and $\Gamma_R$ are the mass
and total width of the resonance, $B_R$ is the branching ratio of its decay
into $\pi\rho$ and $q_{cm}$ is the center of mass momentum
\be
q_{cm} =\sqrt{[s-(m_{\rho}+m_{\pi})^2]
[s-(m_{\rho}-m_{\pi})^2]}/2\sqrt{s} \, .
\label{qcm}
\ee
Here $F_s$ is the spin factor and $F_i$ is the isospin factor.
The latter is equal to $(1/2)\times (2/3)=1/3$
for $I_R=1$. The first factor reflects the fact that we are interested
in $\rho^0 \pi$ scattering, and only one of two decay channels of an
$I_R =1$ resonance can contribute here, $R^{\pm} \to \rho^0 \pi^{\pm}$
but not $\rho^{\pm} \pi^0$. The second factor corresponds to the
assumption that all three pion isospin states are equally populated
in the gas.  Similarly, for $I_R=0$ the isospin factor is
$(1/3)\times (1/3)=1/9$.  We take into account the following resonances [9]:
 $a_1(1260)$, $\pi(1300)$, $a_2(1320)$ and $\omega(1420)$.
The nearest resonance under the threshold, $\omega(782)$, contributes
a negligilbe amount due to its narrow width.
For the spin factor we take $F_s= 1,~1,~2,~1$,
respectively, for the aforementioned resonances. (These factors correspond
to transverse $\rho$-mesons only.) The amplitude in the
pion rest frame is obtained from Eq. (4) by multiplication by the rescaling
factor $k_{\rho}/q_{cm} = \sqrt{s}/m_{\pi}$, where $k_{\rho}=
\sqrt{E_{\rho}^2-m_{\rho}^2}$ is the $\rho$ momentum in the pion rest frame.

For $\sigma_{\rho\pi}$ we use the standard resonance formula.
\be
\sigma_{\rho \pi} =
\sum_R~F_sF_i\frac{\pi}{q^2_{cm}}\frac{B_R\Gamma^2_R}{(\sqrt{s} -m)^2 +
\Gamma^2_R/4} \, .
\ee
According to Adler's theorem the pion scattering amplitude on any
hadronic target vanishes at zero pion energy in the target rest frame
in the limit of massless pions. In the framework of an effective
Lagrangian this can be achieved if the pion field enters by its derivative
$\partial  \varphi/\partial x_{\mu}$. We assume that in $\rho\pi$
scattering through the $a_1$ resonance $\partial \varphi/\partial x_{\mu}$ is
multiplied by the $\rho$-meson field strength tensor $F_{\mu\nu}$
and the $a_{1\nu}$ field.  This results in the appearance of an
additional factor in ${\rm Re}f_{\rho\pi}$ and in $\sigma_{\rho\pi}$ 
in comparison to Eqs. (4) and (6)
\be
\Biggl ( \frac{s - m^2_{\rho} - m^2_{\pi}}{m^2_{a_1} - m^2_{\rho} -
m^2_{\pi}}\Biggr )^2 \, .
\ee
Here normalization at $s=m^2_{a_1}$  was performed. When  $s > m^2_{a_1}$
this factor is replaced by unity. The analogous factors were also
introduced for other resonance contributions.

At high energies we assume that the Regge approach is valid for $\gamma
\pi$  scattering and apply the vector dominance model (VDM) to relate $\rho
\pi$ and $\gamma \pi$ amplitudes. As is well known the Regge pole
contributions to the forward scattering amplitude, normalized according to
Eq. (2), have the form:
\be
f(s) = -\frac{k}{4\pi s}\sum_i
~\frac{1+e^{-i\pi\alpha_i}}{\sin \pi\alpha_i}s^{\alpha_{i}}r_i \, ,
\ee
where $\alpha_i$  is the intercept of the $i$'th Regge pole trajectory,
$r_i$ is its residue, and $k$ is the projectile momentum in the target rest
frame. As follows from Eqs. (2) and (8),
\be
\sigma(s) = \sum_i ~r_i s^{\alpha_i-1} \, ,
\ee
\be
{\rm Re}f(s) = -\frac{k}{4\pi s}\sum_i~\frac{1+\cos \pi \alpha_i}{\sin \pi
\alpha_i}r_is^{\alpha_i} \, .
\ee
For $\sigma_{\gamma \pi}$  only $P$ (Pomeron) and $P^{\prime}$ Regge poles
contribute [11,12]. The residues of the $P$ and $P^{\prime}$ poles in
$\gamma\pi$  scattering were found by Boreskov, Kaidalov and Ponomarev (BKP)
[12] using Regge pole factorization and data on $\gamma p$,
$\pi p$  and $pp$ scattering. Taking BKP values of $P$ and $P^{\prime}$
residues we have
\be
\sigma_{\pi \gamma}(s) = 7.48 \alpha ~\Biggl [
\Biggl (\frac{s}{s_0}\Biggr  )^{\alpha_P-1} + 0.971
\Biggl (\frac{s}{s_0}\Biggr  )^{\alpha_{P^{\prime}}-1} \Biggr ] \, ,
\ee
where $\alpha_P=1.0808,~\alpha_{P^{\prime}}=0.5475$, $\alpha=1/137$,
$s_0=1$ GeV$^2$ and $\sigma$ in Eq. (11) is given in millibarns.
For $P$ and $P^{\prime}$
intercepts we take Donnachie-Landshoff values [13]. Since in their fit of
the data BKP assumed $\alpha_P=1$, $\alpha_{P^{\prime}}=1/2$, the values of
the residues in Eq. (11) are slightly changed in comparison with [12]
in order to give the same value of $\sigma_{\pi\gamma}$ at $s=9$
GeV$^2$.  From Eqs. (10) and (11) the
real part of the forward $\gamma \pi$ scattering amplitude can be found:
\be
{\rm Re}f_{\pi \gamma}(s)_{\pi {\rm \, rest \, frame}}
 = -\frac{k}{4\pi} 7.48\alpha \left[
-0.106\Biggl ( \frac{s}{s_0}\Biggr )^{\alpha_P-1} + 0.752
\Biggl ( \frac{s}{s_0}\Biggr )^{\alpha_{P^{\prime}}-1}\right] \, ,
\ee
where the momentum $k$ is in GeV and ${\rm Re}f$ is given in mb$\cdot$GeV.

In VDM  $\sigma_{\rho\pi}(s)$ is related to $\sigma_{\gamma \pi}(s)$ by
[2]
\be
\sigma_{\rho \pi}(s) = \frac{g^2_{\rho}}{4\pi \alpha} \Biggl (1 +
\frac{g^2_{\rho}}{g^2_{\omega}} \Biggr )^{-1} \sigma_{\gamma \pi}(s),
\ee
where $g^2_{\rho}/4\pi =2.54$, $g^2_{\rho}/g^2_{\omega}=1/8$, and the
$\varphi$-meson contribution is neglected. A similar relation holds
for ${\rm Re}f_{\rho \pi}$.   Unlike [2], we prefer here to use direct Regge
formulae for Re$f$ at high energies instead of inferring it from $\sigma$ by
the dispersion relation since, in the latter
approach\footnote{We use this occasion to correct the misprint in the
corresponding equation in Ref. [2] -- in the equation (7) of [2] instead of the
factor $1/(2 \pi)^2$ it should be $1/2 \pi^2$. In the calculations of [2] in
fact the correct factor was used.}, the results are sensitive to the
low energy domain, which is more uncertain.

The  results of the calculations of $\sigma_{\rho\pi}$and Re$f_{\rho\pi}$ as
functions of $\rho$-meson energy in the pion rest frame are presented
in Fig. 1.  As may be seen from the figure the matching of low and
high energy curves is satisfactory.

For the amplitude ${\rm Re} f_{\rho N}$ at laboratory energies of the $\rho$
above 2 GeV we use the results of Ref. \cite{ei} obtained with the dispersion
relation, VDM and experimental data on $\sigma_{\gamma N}$.
At lower energies we again use the resonance approximation
\be
{\rm Re} f_{\rho N}(s)=-\frac{1}{4}~ \frac{1}{2q_{cm}}\sum_{R}(2J_R+1)
F_i
\frac{\Gamma_R^{\rho N} (\sqrt{s}-m_R)}{(\sqrt{s}-m_R)^2 +\Gamma_R^2/4} \, .
\ee
The factor of 1/4 appears because we consider only  transverse
$\rho$-mesons.  The isospin factors $F_i$ are 1/3 and 2/3, respectively,
for $N$  and $\Delta$ resonances.
We take 10 $N$ and $\Delta$ resonances with significant branchings into
$\rho N$ and with masses above the
$\rho N$ threshold  and below 2200 MeV as quoted in [9].
This set of baryonic resonances is close to the set used in
[14]. The main difference in comparison with [14] is that the
effective widths $\Gamma^{{\rm eff}}_{\rho N}=
\Gamma^{{\rm R}}_{\rho N}(q_{{\rm cm}}/q^{{\rm R}}_{{\rm cm}})^{2l+1}$
were introduced only for the resonances close to the $\rho N$
threshold ($q^{{\rm R}}_{{\rm cm}}$ is the value of the c.m.
momentum at the resonance). When $q_{{\rm cm}}> q^{{\rm R}}_{{\rm cm}}$
we put
$\Gamma^{{\rm eff}}_{\rho N}=\Gamma^{{\rm R}}_{\rho N}$.
Besides these resonances, two others with masses below the $\rho N$
threshold were accounted for: the $\Delta(1238)$ and the $N(1500)$.
It was assumed that VDM is valid for the contribution of these
resonances to the widths
$\Gamma_{\rho N}$ and $\Gamma_{\gamma N}$ in the following form.
Since both resonances are close to $\rho N$ threshold, we
can write for each of them $\Gamma_{\rho N}=q_{{\rm cm}}\gamma_{\rho N}$
and $\Gamma_{\gamma N}=k_{{\rm cm}}\gamma_{\gamma N}$,
where $q_{{\rm cm}}$ and $k_{{\rm cm}}$ are the $\rho N$ and $\gamma N$
momenta in the c.m., respectively. Then we assume that
$\gamma_{\rho N}$ and $\gamma_{\gamma N}$ are related by the VDM formula
\be
\gamma_{\gamma N} = 4\pi \alpha \frac{1}{g^2_{\rho}}\Biggl ( 1 +
\frac{g^2_{\rho}}{g^2_{\omega}} \Biggr ) \gamma_{\rho N} \, .
\ee
The value of $\gamma_{\gamma N}$ can be found from the values of
$\sigma_{\gamma N}$ at the resonance peaks. The contribution of the
$\Delta(1238)$ and of the $N(1500)$ to Re$f_{\rho N}$ are essential at low
energies: they contribute about $-1$ to $-0.5$ fm at
$E_{\rho}=1 - 2$ GeV in the nucleon rest frame.

The results for
$\sigma_{\rho N}$ and ${\rm Re} f_{\rho N}$ in the rest frame of the nucleon,
the curve obtained in Ref.\cite{ei} for high energies, and the
matching curve are shown in Fig. 2. As can be seen the matching of low
energy and high energy curves is good.

\section{Determination  of $\rho$-meson mass shift and width broadening
based on the nucleon and pion distributions produced
in heavy ion collisions.}

As mentioned above, in heavy ion collisions only nucleons and pions are
considered as constituents of the medium. Therefore, in this case
Eqs. (1) and (3) take the form
\be
\Delta m(E) = -\frac{2\pi}{m}\left[ \rho_N{\rm Re}f_{\rho
N}(E)+\rho_{\pi}{\rm Re}f_{\rho\pi}(E)\right] \, ,
\ee
\be
\Delta\Gamma(E) = \frac{k}{m}\left[ \rho_N
\sigma_{\rho N}(E)+\rho_{\pi}\sigma_{\rho\pi}(E)\right] \, ,
\ee
where $\rho_N$  and $\rho_{\pi}$ are the nucleon and pion densities
during the final stage of evolution of the hadronic matter produced
in heavy ion collisions.

We will restrict ourselves to consideration of central, head-on, collisions
with small values of impact parameter when the number of participants --
the nucleons, which undergo significant momentum transfer -- is close to
the total number of colliding nucleons.

As shown by the experimental data, the nucleons and pions produced in
heavy ion collisions cannot be considered as a gas in global thermal
equilibrium even during the last
stage of evolution of hadronic matter created in the collisions.
In order to demonstate this
let us discuss separately the cases of high energy, $E \sim 100$ GeV$\cdot$A,
and low energy, $E\sim 1-10$ GeV$\cdot$A, heavy ion collisions. In the case of
high energy collisions the longitudinal and
transverse momenta of nucleons and pions are very different. In the
experiment on $S + S$  collisions at 200 GeV$\cdot$A [15] it was
found that $\langle p^{\rm cm}_{LN}\rangle =3.3$ GeV,
$\langle p_{TN}\rangle = 0.61$ GeV, and $\langle p^{\rm cm}_{L\pi}\rangle
\approx 0.70$ GeV, $\langle p_{T\pi}\rangle
\approx 0.36$ GeV.  In other experiments on high energy heavy ion collisions
-- see  [16,17] -- the situation is qualitatively similar.
This means that one can by no means speak about a thermal gas
of final particles in global equilibrium, and their momentum distributions
must be taken from experiment.  This, however, still leaves open the
possibility of local thermal equilibrium.

The data for low energy heavy ion collisions also indicate that pions and
nucleons cannot be described as gases in global thermal equilibrium.
The  angular distributions of pions produced in $Ni+Ni$  collisions at
$E=1-2$ GeV$\cdot$A shows essential anisotropy [18]. If the pion
angular distribution in the centre of mass system is approximated by $1+
a\cos^2 \theta$ then, from the data, follows $a \approx 1.3$. Unfortunately,
there is not enough experimental information on nucleon angular and
momentum  distributions. We have checked the hypothesis of global thermal
equilibrium by assuming that the probability  of production of the given
number of particles is proportional to the statistical weight of the final
state (Fermi-Pomeranchuk approach [19,20]). It is evident that this
hypothesis is even more general than the hypothesis of global thermal
equilibrium.  In this approach the pion/nucleon ratio $R_{\pi}=n_{\pi}/N$ in
central collisions can be predicted in terms of the main ingredient of the
method -- the volume per particle at the final stage of evolution and, of
course, the initial energy.  A calculation shows that the data [18]
on the energy dependence of $R_{\pi}$ are well described by
the statistical model, but in order to get the absolute values of
$R_{\pi}$ in $Ni+Ni$  as well as in $Au+Au$ [21] collisions it is
necessary to put the volume per nucleon very small, about 5 times smaller
than in a normal nucleus, which is unacceptable.
Therefore, the only way to perform the averaging over momentum distributions
of pions and nucleons is by taking the latter from experimental data on heavy
ion collisions.

When calculating the $\rho$-meson mass shift and
width broadening an averaging must be performed over the $\rho$-meson
direction of flight relative to nucleons and pions. Such a calculation can
be done only for real experimental conditions. For this reason we restrict
ourselves to rough estimates.

Consider first the case of high energies.  As an example take the
experiment [15] for central collisions where the ratio of
pion to nucleon multiplicities was found to be $R_{\pi}=5.3$.  Suppose
that in this experiment the $\rho$-meson is produced with longitudinal and
transverse momenta in the laboratory system $k_{L}=3$ GeV,
$k_{T}=0.5$ GeV. We choose these values as typical for such an experiment.
For these values of $\rho$ momenta the formation time of the $\rho$-meson
is close to the mean formation time of pions so a necessary condition
of our approach is fulfilled. Since the mean
momenta of nucleons and pions in the experiment [15] are known (they
were presented above) it is possible, using the curves of Figs. 1 and 2,
to calculate the mean values of Re$f_{\rho N}$, Re$f_{\rho\pi}$,
$\sigma_{\rho N}$ and $\sigma_{\rho\pi}$ in $\rho N$ and $\rho \pi$
scattering. The results are, in lab frame:
\be
\langle {\rm Re}f_{\rho N} \rangle \approx -1.1~{\rm fm} \, ,
~~~~~~~\langle {\rm Re}f_{\rho\pi} \rangle \approx 0.03~{\rm fm} \, ,
\ee
\be
\langle \sigma_{\rho N} \rangle \approx 45 ~{\rm mb} \, ,
~~~~~~~\langle \sigma_{\rho\pi} \rangle \approx 20~{\rm mb} \, .
\ee
The small value of $\langle{\rm Re}f_{\rho\pi}\rangle$ arises from a
compensation of positive and negative contributions from low and 
high energy collisions,
that is, from the scattering of the $\rho$-meson on pions moving in
the same direction (comovers) or in the opposite one. 
Because of this compensation
$\langle{\rm Re}f_{\rho\pi}\rangle$ is badly determined, but since
it is small this fact does not influence the final result.

Using Eqs. (18) and (19) we can now find the mass shift and width
broadening of the $\rho$-meson. For nucleon and pion densities we take
\be
\rho_N = \frac{N}{V} = \frac{N}{N\upsilon_N + n \upsilon_{\pi}} = \frac{1}
{\upsilon_N(1+R_{\pi}\frac{\upsilon_{\pi}}{\upsilon_N})} \, ,
\ee
\be
\rho_{\pi} = \frac{n_{\pi}}{V}= \frac{n}{N\upsilon_N + n \upsilon_{\pi}} =
\frac{R_{\pi}}{\upsilon_N(1 + R_{\pi}\frac{\upsilon_{\pi}}{\upsilon_N})}\, ,
\ee
where $N$ and $n$ are the numbers of nucleons and pions at the last stage
of evolution, $R_{\pi}=n/N$, and $V$ is the volume of system at this stage.
It is assumed that at this stage of evolution any participant -- nucleon or
pion -- occupies the volume $\upsilon_N$ or $\upsilon_{\pi}$, respectively.
We can write
\be
\rho_N= \frac{\rho^0_N}{1 + R_{\pi}\beta}\, , ~~~~\rho_{\pi} =
\frac{\rho^0_NR_{\pi}}{1+R_{\pi}\beta}\, ,
\ee
where $\rho^0_N=1/\upsilon_N$ and $\beta=\upsilon_{\pi}/\upsilon_N$.
For numerical estimates we take
$\rho^0_N=0.3$ fm$^{-3}$, about two times standard nucleon density. This
number is probably one of the most uncertain ingredients of our
calculations.  Substitution of Eqs. (18), (19) and (22) in (16) and (17),
together with the experimental values $R_{\pi} = 5.3$ and $\beta=1$, gives
\be
\Delta m_{\rho} = 18 - 2 = 16~{\rm MeV} \, ,
\ee
\be
\Delta \Gamma_{\rho}\approx 150 + 400 = 550~{\rm MeV} \, .
\ee
The first numbers above refer to the contributions from $\rho-N$ and
second from $\rho-\pi$  scattering. Because the $\rho$-meson width broadening
appears to be very large, a basic condition of our approach, $\Delta \Gamma
\ll m_{\rho}$, is badly fulfilled. The applicability condition of
the method, $\mid{\rm Re}f\mid < d$, is not well satisfied either
since in this case $d=0.9$ fm. For these reasons the values of
$\Delta m_{\rho}$ and $\Delta\Gamma_{\rho}$ may be considered only as 
estimates.

The main conclusion to be drawn from Eqs. (23) and (24) is that for
$\rho$-mesons produced in high energy heavy ion collisions with the above
chosen values of longitudinal and transverse momenta, the mass shift
is small, but the width broadening is so large that one can
hardly observe a $\rho$-peak in $e^+e^-$ or $\mu^+\mu^-$ mass distributions.
Let us estimate how sensitive the results are to variations of $k_L$
and $k_T$. It can easily be seen that the mass shift will always be small,
say $\Delta m_{\rho} \la 50$ MeV. If we put $k_{T}=0$  instead of
$k_{T}=0.5$ GeV, this will only weakly influence the mean value of
$\sigma_{\rho N}$ and decrease $\sigma_{\rho\pi}$ by 20\%. The latter
results in a decrease of $\Delta \Gamma_{\rho}$ by 80 MeV, 
which is within the limits
of accuracy of our estimates. The variation of $k_{L}$ in the range
1 GeV to 10 GeV also results in variations of 10-20\% in $\Delta\Gamma_{\rho}$.

As mentioned above, the main uncertainty in our approach comes from
the assumed value of the nucleon density at the final stage of
evolution: $\rho^0_N =0.3$ fm$^{-3}$. If this density would
be a factor of two smaller then $\Delta\Gamma_{\rho}\sim 250$ MeV and
the $\rho$-meson could be observed as a broad
peak in the $e^+e^-$  or $\mu^+\mu^-$ mass spectrum. It should be mentioned,
however, that the chosen value of $\beta=\upsilon_{\pi}/\upsilon_N=1$
is rather uncertain. If we assume that
$\beta=(r_{\pi}/r_N)^3$, where $r_{\pi}$ and $r_N$  are pion and nucleon
electromagnetic radii, $r_{\pi}=0.66$ fm, $r_N=0.81$ fm, then $\beta\approx
0.55$. The choice of such $\beta$ increases $\Delta\Gamma$  by the factor of
1.6.

In the course of $\rho$-meson propagation in the medium its decay width
$\Gamma(\rho\to\pi\pi)$ may decrease\footnote{One of the
authors -- B.I. -- is grateful to G.Brown for this remark.}.
This effect can be estimated by
substitution of an effective pion propagator in the medium [~$k^2-(m_{\pi} -
i \Gamma_{\pi}/2)^2~]^{-1}$ into the imaginary part of pion loop
determining $\rho \to \pi\pi$ decay. (Here $\Gamma_{\pi}$ is the effective
pion width in the medium arising from pion interaction with medium
constituents.)  The calculation performed in this way gives
\be
\frac{\Gamma(\rho\to \pi \pi)_{\rm medium}}{\Gamma(\rho\to \pi
 \pi)_{\rm vacuum}} =
1 - \frac{3}{8}\Biggl ( \frac{\Gamma_{\pi}}{m_{\rho}}\Biggr )^4 \, .
\ee
Even when $\Gamma_{\pi} \approx 500$ MeV  this
correction is small.

Our qualitative conclusion is that in central collisions of heavy nuclei at
high energies, $E\sim 100$ GeV$\cdot$A, where a large number of pions per
participating nucleon is produced, the $\rho$-peak will be
observed in $e^+e^-$ or $\mu^+\mu^-$ mass distributions only as a
very broad enhancement, or even no enhancement at all.  Inspite of
the assumptions we made, including noninteracting nucleon and pion
matter at the final stage of evolution and the specific numerical
value of the nucleon density, we believe that this qualitative
conclusion is still valid.  This conclusion
is in qualitative agreement with the measurement of $e^+e^-$  pair
production in heavy ion collisions [22] where no $\rho$-peak was found and
only a smooth $e^+e^-$ mass spectrum from 0 to 1 GeV was observed.
If, however, such a peak would be observed in future experiments
it would indicate that the hadronic (nucleon and pion) density at the
final stage of evolution, where the $\rho$-meson is formed, is very low,
even lower than normal nuclear density.

Recently preliminary data in $Pb-Au$ collisions at 160 GeV$\cdot$A
have been presented [23] where, in studying the
$e^+e^-$ mass spectrum, it was
found that the $\rho$-peak is absent at $k_T(e^+e^-)< 400$ MeV,
but reappears as a broad enhancement at $k_T(e^+e^-)> 400$ MeV.
We do not see the possibility for such a phenomenon in the
framework of our approach for central heavy ion collisions.
Moreover, we believe that for central
collisions the absence of a $\rho$-peak at low $k_T$ and its reappearance at
higher $k_T$  will be hard to explain in any reasonable model. The only
explanation we see for this effect is that in this
experiment peripheral $\rho$-meson production plays an essential role.
Then $\rho$-mesons with higher $k_T$ have a larger probability to escape the
collision region and decay as free ones.

Let us turn now to the case of lower energy heavy ion collisions,
$E \sim$ a few GeV$\cdot$A. Consider, as an example, heavy ion collisions at
$E_{\rm kin}=3$ GeV$\cdot$A and production of $\rho$-mesons of energy
$E^{\rm tot}_{\rho}=1.2$ GeV in the forward direction. (This particular
value of the $\rho$-meson energy was chosen because our approach works
better at higher $E_{\rho}$, and $\rho$-mesons of this energy can be
kinematically produced at such a beam energy).
The number of pions produced can be estimated by extrapolation of the data
[18] on $Ni+Ni$  collisions.  This data shows, with good accuracy,
that $R_{\pi}$ is linear in $\sqrt{s}/2 -m$.
We find that $R_{\pi}=0.48$.  As follows from
analysis of the data [18] at $E_{\rm kin}=1.93$ GeV$\cdot$A, the average
energies of produced pions are rather small: $E_{\pi}\sim 200-300$ MeV. At
such low energies it is reasonable to suppose that for pions
$\langle p_L \rangle= \langle p_{\perp}\rangle
\approx 0.2$ GeV.  Assuming that the mean perpendicular momentum of nucleon
participants is the same as at high energy --
$\langle p_{T N} \rangle=0.61$ GeV [15] 
(this assumption does not much influence the final results)
we can construct the momentum distributions of nucleons.
Then we are in a position to calculate the mean values of
Re$f_{\rho N}$, Re$f_{\rho \pi}$, $\sigma_{\rho N}$ and $\sigma_{\rho\pi}$.
The results are:
\be
\langle {\rm Re}f_{\rho N} \rangle= -0.54~{\rm fm} \, ,
 ~~~~~~~~~\langle {\rm Re}f_{\rho \pi} \rangle= 0.30~{\rm fm} \,,
\ee
\be
\langle \sigma_{\rho N}\rangle = 45~{\rm mb} \, ,
~~~~~~~~~~\langle \sigma_{\rho \pi} \rangle= 13~{\rm mb} \, .
\ee
For the $\rho$-meson mass shift and width broadening we have,
with the same value of $\rho^0_N$ as above and $\beta=1$:
\be
\Delta m_{\rho} = 37 - 10 = 27~{\rm MeV} \, ,
\ee
\be
\Delta~\Gamma_{\rho} = 245 + 35 = 280~{\rm MeV} \, .
\ee
The first numbers above refer to $\rho N$ scattering, the second
ones to $\rho\pi$. The conclusion is that in low energy heavy ion
collisions a $\rho$-peak may be observed in $e^+e^-$ or $\mu^+\mu^-$
mass distributions as a broad enhancement approximately at the position of
$\rho$-mass.

\section*{Acknowledgements.}

We are very indebted to K. Boreskov, A. Kaidalov, G. Brown and A. Sibirtsev
for illuminating discussions.
We are thankful to A. Smirnitsky and V. Smolyankin for
the help in getting information about experimental data.
This work was supported by INTAS Grant 93-0283, CRDF grant RP2-132,
Schweizerischer National Fonds grant 7SUPJ048716, RFBR grant
97-02-16131, and U.S. Department of Energy grant DE-FG02-87ER40328.
V. L. E. acknowledges support of BMBF, Bonn, Germany.

\newpage

\section*{Figure Captions}

\noindent
Figure 1: Cross section (a) and real part of the forward scattering amplitude
(b) for $\rho$-mesons scattering on pions as functions of the total  
$\rho$-meson energy in the pion rest frame. The curves at low energy
are the result of the resonance approximation.  
The curves at high energy
are the result of the Regge parametrization.  These curves
are matched at intermediate energies.\\

\noindent
Figure 2: Same as for Fig. 1 but for $\rho$-mesons scattering on
nucleons.  The curves at high energy are from Ref. [2].

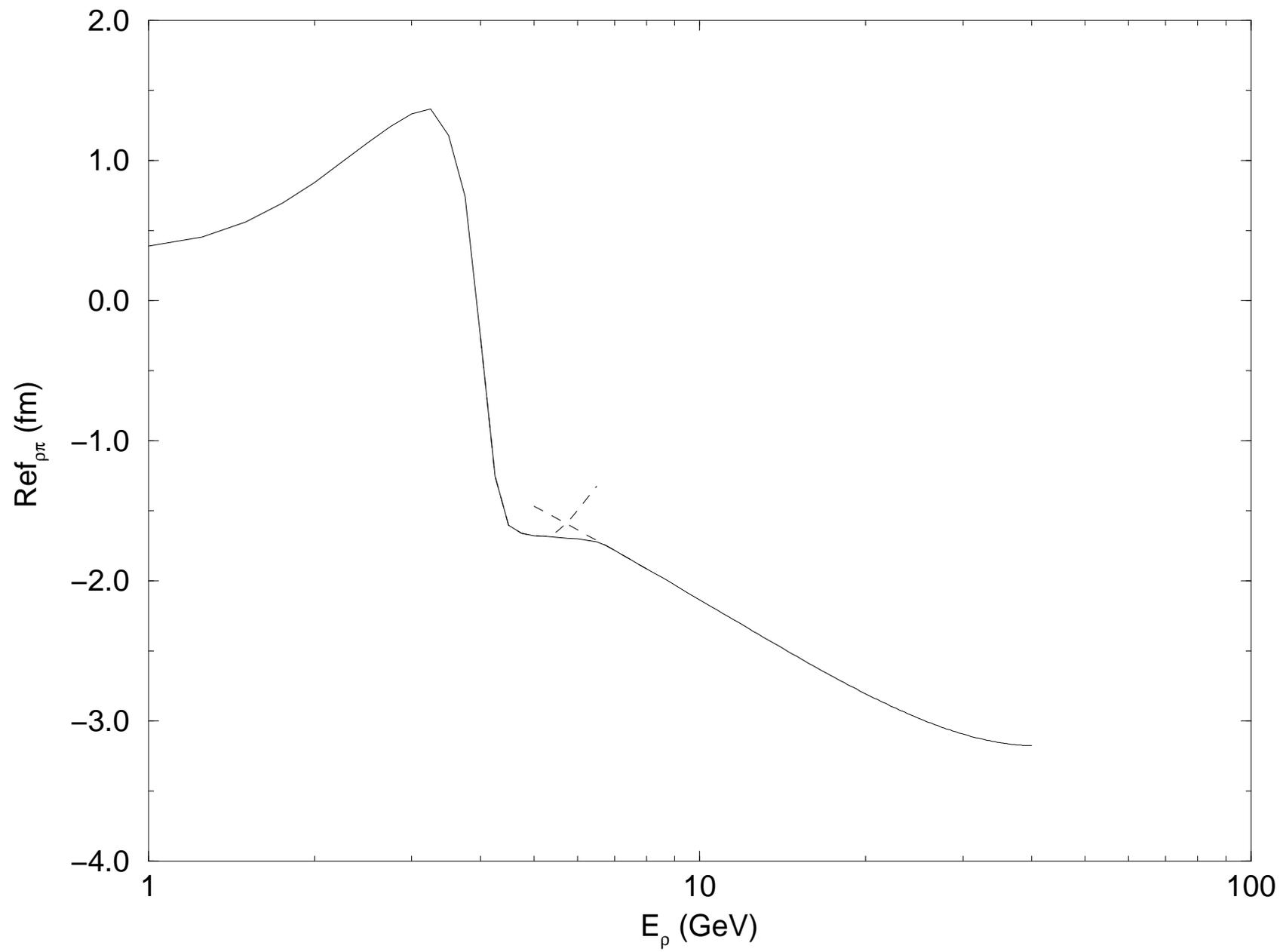
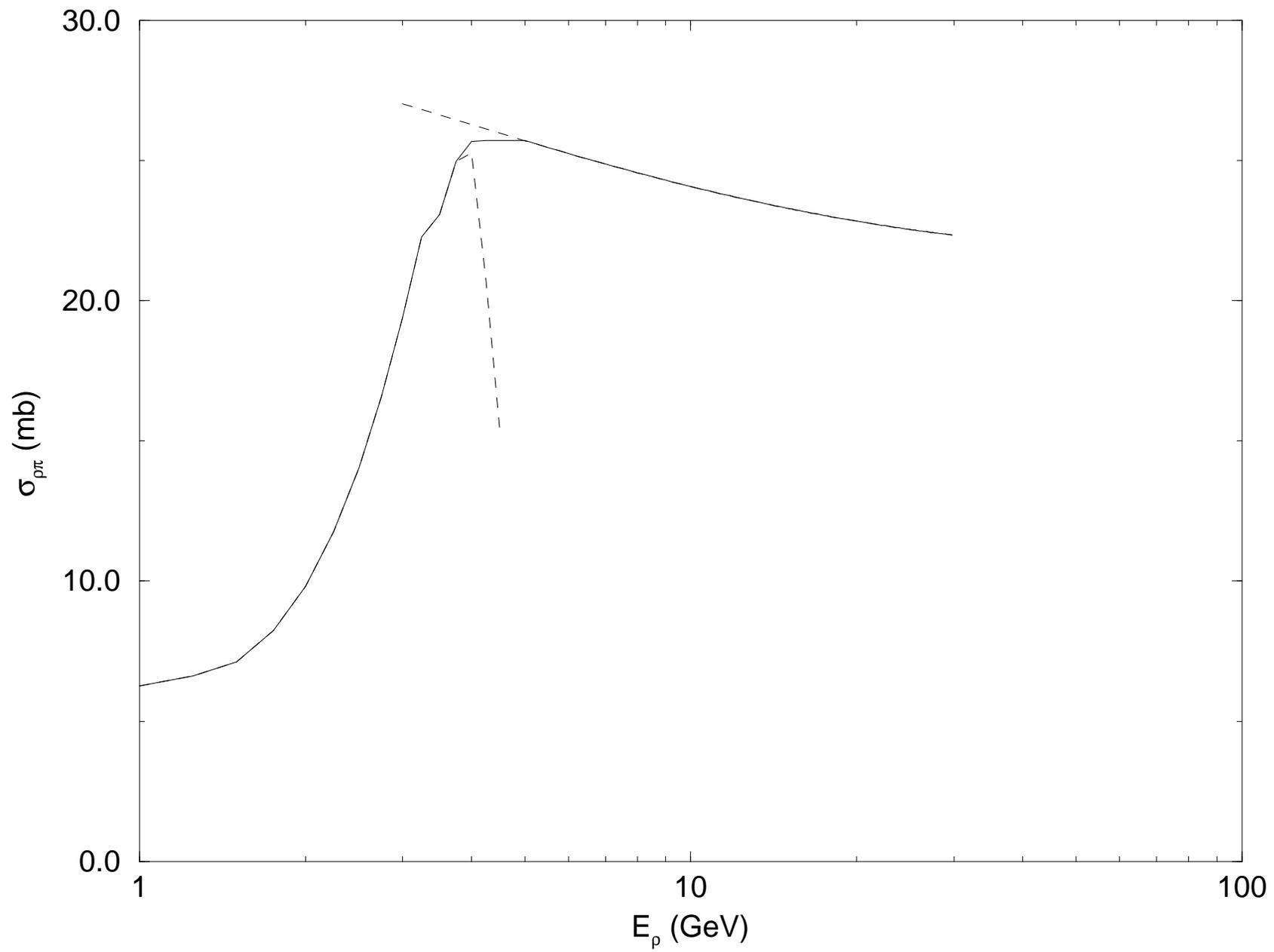
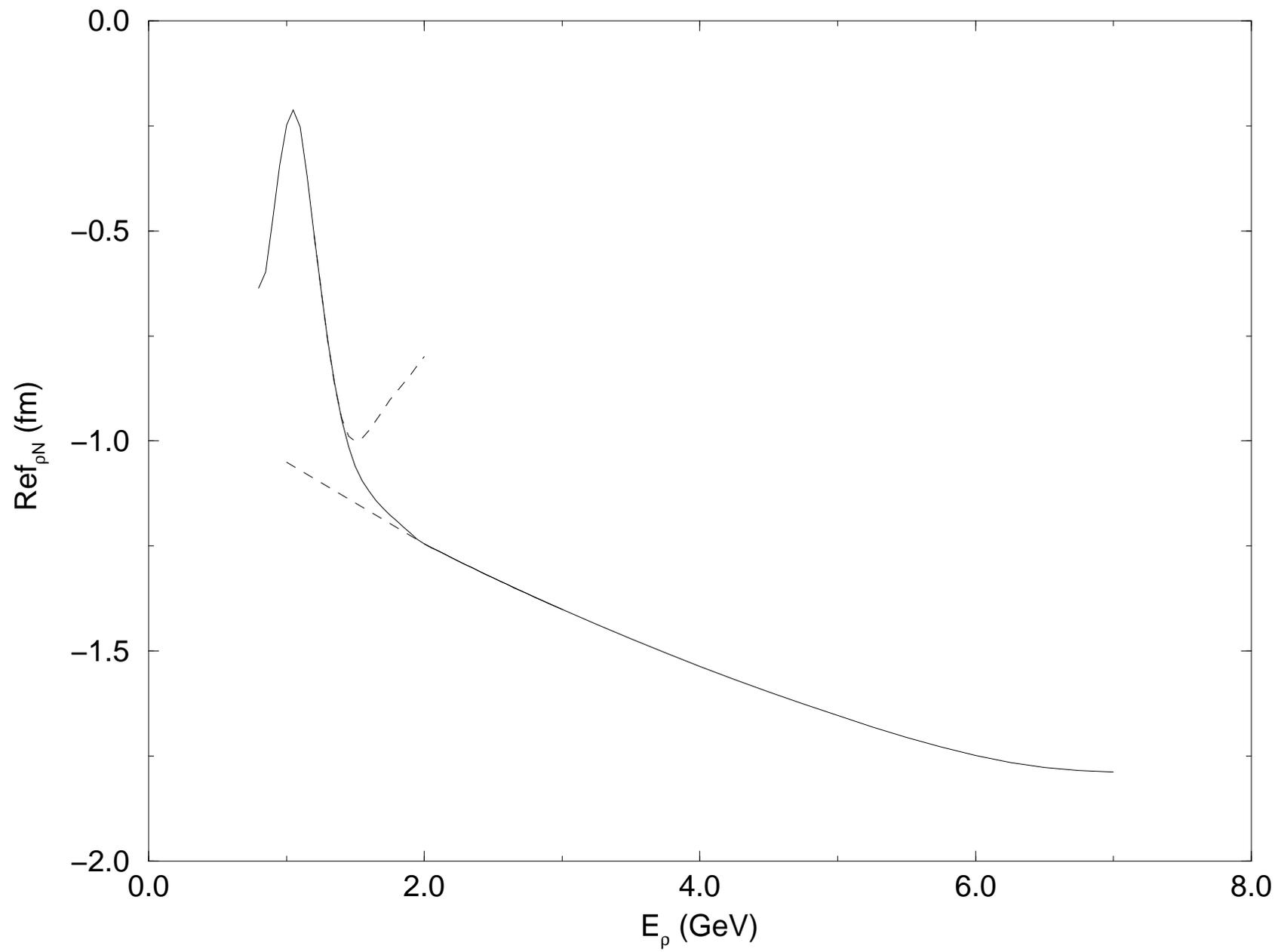
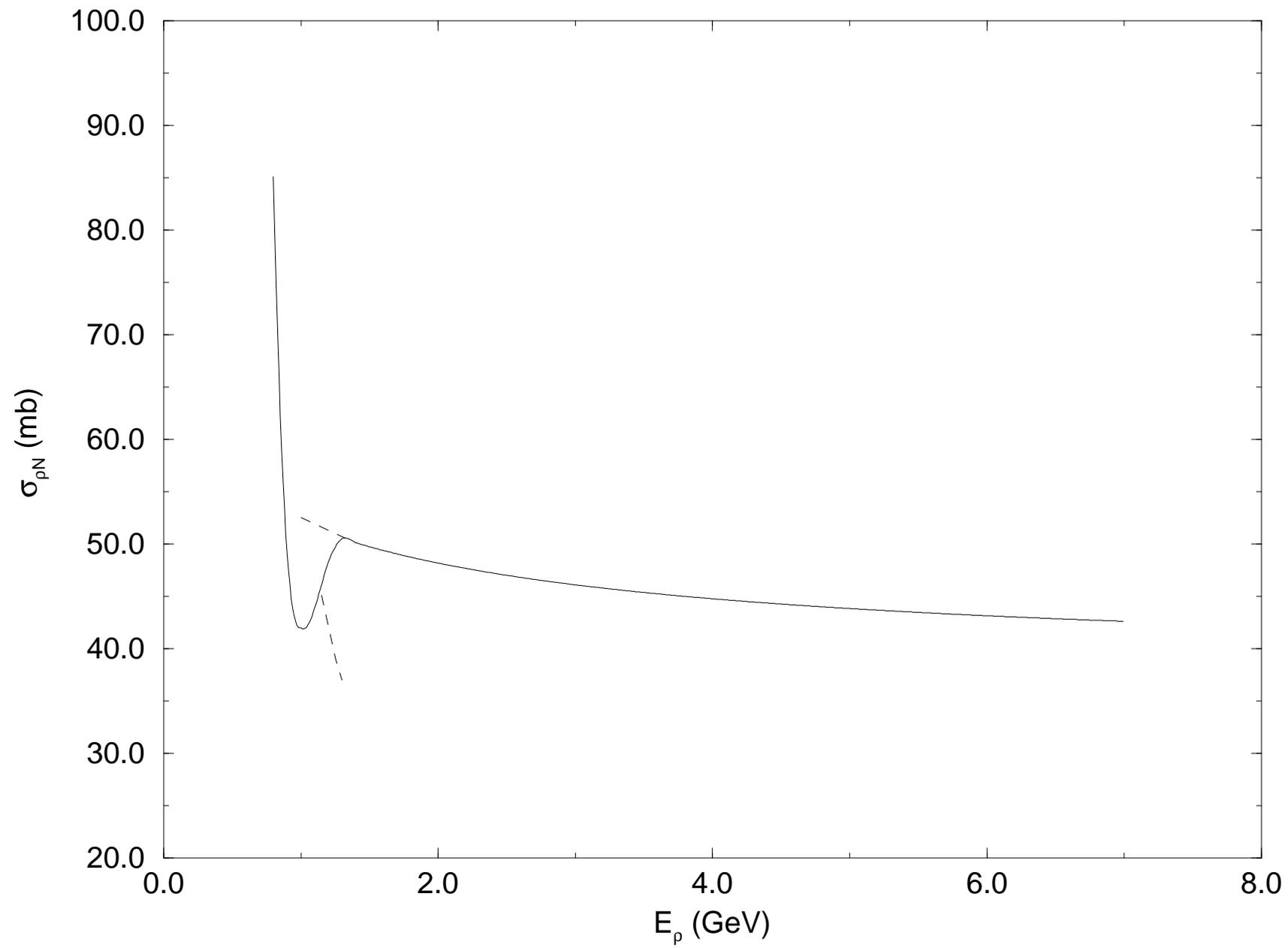
\end{document}